\documentclass[iop]{emulateapj}

\def\sun{\odot}
\def\earth{\oplus}
\def\icarus{Icarus}

\usepackage{epsfig}
\begin{document}
\title{The Last Stages of Terrestrial Planet Formation: Dynamical Friction and
the Late Veneer} \shortauthors{Schlichting} \shorttitle{The Last Stages of
Terrestrial Planet Formation} \author{Hilke E. Schlichting\altaffilmark{1,2,3},
Paul H. Warren\altaffilmark{1}, Qing-Zhu Yin\altaffilmark{4}}\altaffiltext{1}
{UCLA, Department of Earth and Space Science, 595 Charles E.  Young Drive
East, Los Angeles, CA 90095}\altaffiltext{2} {Department of Astronomy and
  Astrophysics, California Institute of
Technology, MC 130-33, Pasadena, CA 91125} \altaffiltext{3} {Hubble Fellow}
\altaffiltext{4}{UCD, Department of Geology, One Shields Avenue, Davis, CA
95616}\email{hilke@ucla.edu}

\shorttitle{The Last Stages of Terrestrial Planet Formation}

\begin{abstract}
The final stage of terrestrial planet formation consists of the cleanup of
residual planetesimals after the giant impact phase. Dynamically, a residual
planetesimal population is needed to damp the high eccentricities and
inclinations of the terrestrial planets to circular and coplanar orbits after
the giant impact stage. Geochemically, highly siderophile element (HSE)
abundance patterns inferred for the terrestrial planets and the Moon suggest
that a total of about $0.01~M_{\earth}$ of chondritic material was delivered
as `late veneer' by planetesimals to the terrestrial planets after the end of
giant impacts. Here we combine these two independent lines of evidence for a
leftover population of planetesimals and show that: (1) a residual population
of small planetesimals containing $0.01~M_{\earth}$ is able to damp the high
eccentricities and inclinations of the terrestrial planets after giant impacts
to their observed values. (2) At the same time, this planetesimal population
can account for the observed relative amounts of late veneer added to the
Earth, Moon and Mars provided that the majority of the accreted late veneer
was delivered by small planetesimals with radii $\lesssim 10~\rm{m}$. These
small planetesimal sizes are required to ensure efficient damping of the
planetesimal's velocity dispersion by mutual collisions, which in turn ensures
sufficiently low relative velocities between the terrestrial planets and the
planetesimals such that the planets' accretion cross sections are
significantly enhanced by gravitational focusing above their geometric
values. Specifically we find, in the limit that the relative velocity between
the terrestrial planets and the planetesimals is significantly less than the
terrestrial planets' escape velocities, that gravitational focusing yields a
mass accretion ratio Earth/Mars $\sim (\rho_{\earth}/\rho_{mars})
(R_{\earth}/R_{mars})^4 \sim 17$, which agrees well with the mass accretion
ratio inferred from HSEs of 12-23. For the Earth-Moon system, we find a mass
accretion ratio of $\sim 200$, which, as we show, is consistent with estimates
of 150-700 derived from HSE abundances that include the lunar crust as
well as mantle component. We conclude that small residual planetesimals
containing about $\sim 1\%$ of the mass of the Earth could provide the
dynamical friction needed to relax the terrestrial planet's eccentricities and
inclinations after giant impacts, and also may have been the dominant source
for the late veneer added to Earth, Moon and Mars.
\end{abstract}


\section{INTRODUCTION}
Terrestrial planet formation is generally considered to consist of three main
 stages. The first stage consists of the formation of small planetesimals
 \citep[e.g.][]{CY10}, the second stage of the coagulation of these small
 planetesimals into roughly Mars-sized protoplanets
 \citep[e.g.][]{IM93,WSDMO97}, and the third stage is comprised of collisions
 of a few dozen protoplanets, called giant impacts
 \citep[e.g.][]{ACL99,C01}. However, several lines of evidence suggest that a
 significant amount of mass was left in planetesimals at the end of the giant
 impact phase in the terrestrial planet region. This therefore argues for an
 additional and final stage of terrestrial planet formation, which consists of
 the clean-up of the leftover planetesimals. This final stage of
 terrestrial planet formation is the focus of the present paper.

Evidence for a significant population of planetesimals in the terrestrial
planet region at the end of the giant impact phase comes from two different
areas of research:

Geochemical evidence suggests that the Earth accreted chondritic materials
equivalent to about 0.3\%-0.7\% of its total mass after the end of giant
impacts \citep{W09}. The evidence for this `late veneer' comes from highly
siderophile elements (HSEs) that are geochemically characterized as having a
strong tendency to partition into metal relative to silicates. Hence the
silicate portions of terrestrial planets with rocky cores are expected to be
effectively stripped of HSEs after final core segregation. However, a
surprisingly high abundance of HSEs has been inferred on the terrestrial
planets, which suggests continued planetesimal accretion onto the Earth, Moon
and Mars after core formation \citep{W99,WH04,W09}. This
implies for the Earth-Moon system that a significant amount of mass was
accreted after the Moon-forming impact. The geochemical evidence for the late
veneer and the estimated amounts of material added to the Earth, Moon and Mars
after the giant impact phase are summarized and discussed in detail
in section 3.

Independently, there is evidence from planet formation models that argues for
a population of leftover planetesimals in the terrestrial planet region (see
section \ref{s1} for details). This evidence comes from planet formation
models that examine the onset of giant impacts \citep{KB06,FC07}, from
simulations of collisions between protoplanets \citep{BA99} and from the
requirement to relax the high eccentricities and inclinations of terrestrial
planets after giant impacts \citep{C01,OBM06,R06,SR06}. These works argue for
the existence of a planetesimal population in the terrestrial planet region
that still contains a few to 10\% of the mass of the Earth following the epoch
of giant impacts.

In this paper, we investigate the accretion of the leftover planetesimals
after the end of giant impacts in the inner solar system and show that
we can account for the relative amounts of late veneer accreted by the Earth,
Moon and Mars provided that most of it was delivered by small
planetesimals. In section 2 we first estimate the mass left in planetesimals
after giant impacts from planet formation models.  We summarize the
geochemical evidence for a late veneer and estimate the mass of late veneer
added to Earth, Moon and Mars in section 3. We calculate the accretion cross
section for the Earth, Moon and Mars for a range of planetesimals velocities
and compare our results with the inferred late veneer for these bodies from
measured HSE abundances in section 4.  We calculate the typical planetesimal
sizes that delivered the late veneer in section 5 and estimate their accretion
timescale in section 6. Discussion and conclusions follow in section 7.

\section{Leftover Planetesimals after the End of Giant-Impacts: Evidence from Planet Formation Models}\label{s1}

Orbit crossing and giant impacts begin once mutual stirring of the
protoplanets can no longer be efficiently damped by small planetesimals. Order
of magnitude estimates that balance the stirring rates of the protoplanets
with the damping rates due to dynamical friction, which is generated by small
planetesimals, find that orbit crossing sets in when $\sigma \sim \Sigma$,
where $\Sigma$ and $\sigma$ correspond to the mass surface density in
protoplanets and small planetesimals, respectively \citep{GLS042}. This result
has been confirmed by numerical simulations studying the onset of this
instability in the terrestrial planet region \citep{KB06}. Therefore about
50\% of the total mass still resides in small planetesimals when giant impacts
begin. During giant impacts planetesimal accretion continues and additional
`new' planetesimals may be produced as byproducts of giant impacts. Therefore,
due to the fact that giant impacts set in when $\sigma \sim \Sigma$ and
the possible production of `new' planetesimals in giant impacts themselves, a
significant population of planetesimals is expected to still be present after
the end of giant impacts.

In addition, leftover planetesimals provide a way to relax
the high eccentricities and inclinations of terrestrial planets after giant
impacts. N-body simulations predict eccentricities and inclinations for
terrestrial planets after giant impacts that are significantly larger than the
time averaged values of the terrestrial planets in our solar system
\citep{CW98,ACL99,C01}. A population of small planetesimals could provide the
dynamical friction that would be needed to damp the eccentricities and
inclinations after giant impacts to the observed values of the terrestrial
planets. It has already been shown in direct N-body integrations that
including a population of less massive planetesimals, in addition to the
massive planetary embryos, decreases the final eccentricities and inclinations
\citep{CW98,C01,OBM06,R06}. However, due to computational limitations, none of
these works were able to include planetesimals small enough such that their
collective interactions could be accurately described by dynamical
friction. In the limiting case where the terrestrial planets are embedded in a
large number of small planetesimals, the damping of the velocity dispersion
(i.e. the damping of the eccentricity and inclination) by dynamical friction
only depends on the total mass surface density of the planetesimals and is
independent of the mass of the individual bodies. Since this limit has not
been reached in direct N-body simulations, such works underestimate the
strength of dynamical friction for a given mass surface density of
planetesimals and hence overestimate the mass needed in planetesimals to damp
the eccentricities and inclinations.  Further, the velocity dispersion of the
small planetesimals, $u$, is likely to be damped by mutual planetesimal
collisions, which in turn may lead to more effective dynamical friction being
exerted on the terrestrial planets. This is because the strength of dynamical
friction depends on the relative velocity between the planetesimals and the
terrestrial planets, $v_{rel}$. Collisional damping of the planetesimal
velocity dispersion has not been modeled by direct N-body simulations, because
including of the order of $\sim 10,000$ small planetesimals is still not
feasible computationally.

The minimum mass in small leftover planetesimals needed to damp the
eccentricities and inclinations of the terrestrial planets to their current
values can be estimated by comparing the eccentricity damping timescale to the
accretion timescale of the leftover planetesimals. The damping timescale,
$t_{damp}$, due to dynamical friction generated by leftover planetesimals is
given by
\begin{equation}
t_{damp}=-v \frac{dt}{dv} \sim \frac{\rho R}{\sigma \Omega}
\left(\frac{v}{v_{esc}}\right)^4
\end{equation}
where $\rho$ is the mean density, $R$ the radius, $v_{esc}$ is the escape
velocity of the terrestrial planets and $v$ their velocity dispersion
\citep{GLS04}; $\Omega=\sqrt{GM_{\sun}/a^3}$ is the Keplerian angular
frequency around the Sun, where $M_{\sun}$ is the mass of the Sun and $a$ the
semi-major axis. The damping timescale needs to be shorter than the time
required for the remaining planetesimals to be accreted by the terrestrial
planets. Writing the
planetesimal accretion timescale as
\begin{equation}
t_{acc}=-\sigma \frac{dt}{d\sigma} \sim \frac{\rho R}{\Sigma
  \Omega}\left(\frac{v}{v_{esc}}\right)^2
\end{equation}
and requiring that $t_{acc} > t_{damp}$ yields
\begin{equation}\label{e3}
\sigma \gtrsim \Sigma \left(\frac{v}{v_{esc}}\right)^2
\end{equation}
where $\Sigma$ corresponds here to the mass surface density of the terrestrial
planets. Hence we see from Equation (\ref{e3}) that we can place a lower limit
on the mass required in small planetesimals as long as we know the velocity
dispersion of the terrestrial planets, $v$, at the end of giant impacts. We
assumed in Equations (1)-(3) that $v>u$, where $u$ is the planetesimal
velocity dispersion, because small planetesimals are likely to damp their
velocities by mutual collisions ensuring that $u \ll v$. The relative velocity
between the planetesimals and terrestrial planets will therefore be dominated
by the terrestrial planet's velocity dispersion. We can estimate a minimum
value for $v$ by requiring that the velocity dispersion must have been at
least large enough for neighboring planetary embryos that are undergoing giant
impacts to cross their orbits. N-body simulations find a typical spacing of
planetary embryos of $\sim 8 R_H$ \citep{L93}, where $R_H$ is the Hill radius
defined as $R_H \equiv a(M/3M_{\Sun})^{1/3}$ where $M=4\pi \rho R^3/3$ is the
mass of the planetary embryo. This suggests that $v \sim 4 \Omega R_H \sim 2.0
(R/a)^{1/2} (M_{\Sun}/M)^{1/6} v_{esc}$, which evaluates to $v \sim 0.1
v_{esc}$ for Earth-like terrestrial planets at 1~AU. Substituting for $v$ in
Equation (\ref{e3}) we find
\begin{equation}\label{e4}
\sigma \gtrsim 0.01 \Sigma .
\end{equation}
This order of magnitude estimate suggests that at least 1\% of the total mass
needs to still reside in small planetesimals after the giant impact phase in
order to damp the terrestrial planets' eccentricities and inclinations to
their observed values. For comparison, N-body simulations of the giant impact
phase, that include smaller planetesimals but that are not in the limit in
which the collective interactions of the small bodies with the terrestrial
planets can be accurately described by dynamical friction, predict
eccentricities of $\sim 0.1$ for terrestrial planets after giant impacts
\citep{C01}. These simulations therefore suggest $v \sim 0.27 v_{esc}$ and
hence that $\sigma \gtrsim 0.07 \Sigma$ \citep{SR06}. Since Equation
(\ref{e4}) yields only a lower limit on $\sigma$ and since it is only an order
of magnitude estimate, we will assume throughout the rest of the paper that at
least 1\% of the total mass needs to reside in small planetesimals after the
giant impact phase in order to damp the terrestrial planets' eccentricities to
their observed values. In addition, we note here that the value derived in
Equation (\ref{e4}) could be reduced by a factor of $\sim (a \Omega/6 v)^4$ if
gaps form in the planetesimal disk (see section 6 for details). This is due to
the fact that the presence of gaps increases the planetesimal accretion
timescale, while not significantly altering the damping timescale.

\section{Geochemical Evidence for Planetesimal Accretion after the Giant
  Impact Phase}

\subsection{Earth}
HSEs are comprised of Re, Au and the six platinum-group elements Os, Ir, Ru,
Pt, Rh and Pd. These elements have very high metal-silicate partition
coefficients, which suggests that the silicate portions of rocky bodies with
metallic cores should have been stripped of HSEs at the end of core formation.
Yet, the relative abundances of these elements in Earth's mantle are
broadly similar to chondrites. Absolute concentrations of Ir and Os in the
Earth's upper mantle are estimated to be $\sim 3-4\rm{ng/g}$
\citep{W09}. Although this is more than 100 times lower than the
concentrations of these elements found in chondrites, which typically range
from $\sim 400-800\rm{ng/g}$, these concentrations are higher than low
pressure partition coefficients predict \citep{H03,W09}. Therefore, if the
entire mantle harbors HSE abundances similar to the estimate for Earth's
upper mantle, then this suggests, for an Earth mantle mass of $\sim 4.0 \times
10^{27}\rm{g}$, that about $1.5-4.0 \times 10^{25}~\rm{g}$, or
$0.3-0.7\%~M_{\earth}$, of chondritic material was added to the Earth by late
accretion.

\subsection{Mars}
The abundance of HSEs estimated for the Martian mantle is roughly similar to
that of Earth's primitive upper mantle \citep{W09}. \citet{W99} estimate
the primitive mantle abundance of Re, Os, Ir and Au for Mars and find that
the Martian HSE abundances likely range from 0.34-0.66 of the terrestrial
values. Assuming that the Martian mantle has a mass of $5.1 \times
10^{26}\rm{g}$, we estimate a mass accretion ratio for Earth/Mars of 12-23,
i.e. the Earth accreted 12 to 23 times more mass as late veneer than Mars. In
this estimate we assumed that the mass accreted by Mars had an HSE
composition similar to that accreted by the Earth.

\subsection{Moon}
The total amount of late veneer added to the Earth can be reasonably well
estimated from the HSE abundances in its mantle but the situation for the Moon
may be somewhat different, because of the rapid formation of a permanent lunar
crust by $\sim 100~\rm{Myr}$ after the Moon-forming impact, which contrasts
with the delayed development of terrestrial cratons
\citep{CL88,SB00,WH04}. Consequently, the abundances of HSEs in the lunar
mantle probe the material accreted before the isolation of the lunar mantle by
the crust, whereas the HSE abundances in the lunar crust provide constraints
on the accretion of material after crust formation.

Estimates of the HSE content in the lunar mantle have varied considerably from
amounts similar to that of Earth's mantle \citep{R92}, to amounts about 20
times lower \citep{WJ89,WH04,D07,W09}. However recent works generally tend to
favor the lower end of this range \citep{WH04,D07}. Using the \citet{W09}
estimate that the lunar mantle has a factor of 20 lower HSE concentrations
than the terrestrial mantle and assuming a lunar mantle mass of $7 \times
10^{25}~\rm{g}$, we find that $1.3 -3.5 \times 10^{22}\rm{g}$ of late veneer
was added to the lunar mantle, yielding an Earth/Moon mass accretion ratio of
about 1100. However, this estimate ignores an important additional HSE
repository: the crust and the upper part of the lunar lithosphere.

\citet{WH04} suggested that a significant amount of late veneer may have been
deposited into the lunar crust rather than mantle, implying that the majority
of the late veneer was accreted by the Moon after the formation of a permanent
lunar crust. Impact melt breccias and bulk regolith samples have Os and Ir
concentrations averaging 5-15~ng/g \citep{MG76,NB02}. Assuming a lunar
crustal mass of $5 \times 10^{24}~\rm{g}$ and that the late veneer was
delivered by bodies with Os and Ir concentrations similar to chondrites, which
typically range from $\sim 400-800\rm{ng/g}$, and that Os and Ir
concentrations are roughly uniform throughout the crust, no more than $0.3 -
1.9 \times 10^{23}\rm{g}$ of late veneer was added to the lunar crust
\citep{WH04} \footnote{Note, \citet{WH04} quote a
slightly narrower range of $4 - 8 \times 10^{22}\rm{g}$ for the late veneer,
because he assumed a slightly narrower range of possible chondritic
concentrations of Os and Ir than we have used here and assumed a lunar crustal
mass of $3.7 \times 10^{24}\rm{g}$.}. This yields an
Earth/lunar crust mass accretion ratio ranging from 200 to 700. In this
estimate we assumed that the Earth and Moon accreted chondritic material with
the same Os and Ir abundances such that the actual abundance value cancels in
the relative mass accretion ratio. The above estimate for the mass added by
late accretion to the Moon assumed, probably somewhat unrealistically, uniform
Os and Ir concentrations throughout the lunar crust; below we derive an
independent estimate of the late veneer added to the Moon by examining the
ejecta layer and the HSE abundance gradient within it.

The top of the lunar lithosphere is a layer of ejecta accumulated from
countless impacts. This layer is commonly termed a megaregolith, but confusion
arises because the term megaregolith implies loose debris. Pressure-sensitive
sintering \citep{WA11} has probably markedly increased cohesion within the
deeper portion of the ejecta layer. For the loose-debris subvolume of the
ejecta, a thickness of roughly 2.5 km has been inferred based on radar
constraints on blockiness of ejecta from large craters \citep{T79,T09}. But
the ejecta-volume model of \citet{WA11} combined with \citet{F11}'s inventory
of 90 likely lunar basins with diameters $\gtrsim 300~\rm{km}$ suggests a
global mean ejecta accumulation of $\sim 5.8~\rm{km}$. Adding the Procellarum
basin to \citet{F11} inventory would increase the accumulation by $\sim
3.5~\rm{km}$. In Frey's judgment Procellarum shows no topographic basin, but
if not a single impact, this giant region of low elevation and thin crust may
reflect a cluster of a few extremely ancient (degraded), large basins not
included in \citet{F11}. Thus, a reasonable compromise basis for estimating
the global mean ejecta accumulation is to add a large fraction, say 1/3, of a
Procellarum equivalence of thickness, i.e., $\sim 1.2~\rm{km}$. This leads to
a final estimate that the `known' global ejecta accumulation thickness is
$\sim 7~\rm{km}$. However, being based on only observable basins $\gtrsim
300~\rm{km}$ in diameter, this estimate is likely low by a significant factor,
of at least order 2.

The global ejecta layer consists mostly of jumbled target (lunar) matter, with
a much smaller proportion of impactor/chondritic matter. For estimating the
bulk composition of the ejecta layer, the most useful samples are from highly
immature regolith. Immature regolith has been thoroughly churned and mixed,
but not (at least, not for long) at the very surface of the Moon, and thus is
free or nearly free of micrometeorite component and associated enrichment in
HSE \citep{MK91,W04}. For the most commonly measured of the HSE, Ir, the
average composition of all highland regolith samples is $\sim 12~\rm{ng/g}$
\citep{HW91}. For immature highland regolith, with no micrometeorite
component, the average is probably more like $8-10~\rm{ng/g}$.

The ejecta layer may be only a fraction of the total upper-lithosphere
component of the Moon's late veneer. Settling of metals in basin-scale impacts
has probably produced local HSE concentrations deep within and even below the
crust. An iron-meteoritic or ordinary chondritic impactor would contain metal
as a major mineral. In large events, where the central, unejected mass of
impact melt is slow to cool and solidify, the dense metal component must tend
to settle to the very bottom of the impact melt volume. A known example of
metal that settled out of a lunar impact melt is the mostly metallic 4.4 gram
rock 14286 \citep{A95}. Assuming equilibration occurs, only a tiny proportion
of metal would suffice to efficiently scavenge the HSE out of a silicate
melt-metal system. Diffusion within metal is very rapid. The limiting factor,
for the efficiency of HSE scavenging, may be a tendency for the metal
components to be so extremely fine grained that they fail to settle. However,
in the largest events the central mass of unejected impact melt is so slow to
cool that even its silicate crystallization is believed to entail
gravitational differentiation \citep{W96,IM10}. The sunken metal probably
ended up mostly near the bottom of the `sheet' of central, unejected impact
melt, at a depth equivalent to roughly 1/10 the diameter of the transient
crater \citep{W96}; i.e., in general, roughly 1/20 the final basin
diameter. In other words, the depth at which the metal components
predominantly settled was probably of order 50~km. It would take a
subsequent basin half as large as the original transient crater, if centered
at precisely the same point, to begin to excavate the base of the melt
`sheet'; i.e., the settled metal. Thus, HSE concentrations found in the
megaregolith at the present surface of the Moon may under-represent, possibly
by a large factor, the total amount of HSE-rich matter accreted as late
veneer.

Although large basin-scale impacts likely played a crucial role in creating
and churning the lunar ejecta layer, they may not have contributed
significantly to the overall lunar HSE budget of the ejecta layer. This is
because, unlike the small and low velocity impactors that collided with the
Moon early on and which are the focus of this paper, the largest impactors
likely collided late in the lunar history and had large impact velocities,
such that only a fraction of the total impact mass was actually accreted by
the Moon. For example, the Nice model \citep{G05} suggests the average
Late Heavy Bombardment velocity was 21-25 km/s. Under such conditions, at the
most common impact angles, only a fraction of the impactor will actually be
accreted by the Moon, whereas for the Earth, with its much higher escape
velocity, the fraction of impactor matter that fails to accrete is
comparatively negligible. Modeling constraints suggest that for rocky
impactors the lunar accretion efficiency, integrated over all impact angles,
is 0.32-0.16 for 21-25 km/s \citep{AS08,O10}.

\begin{figure}[htp]
\centerline{
\epsfig{file=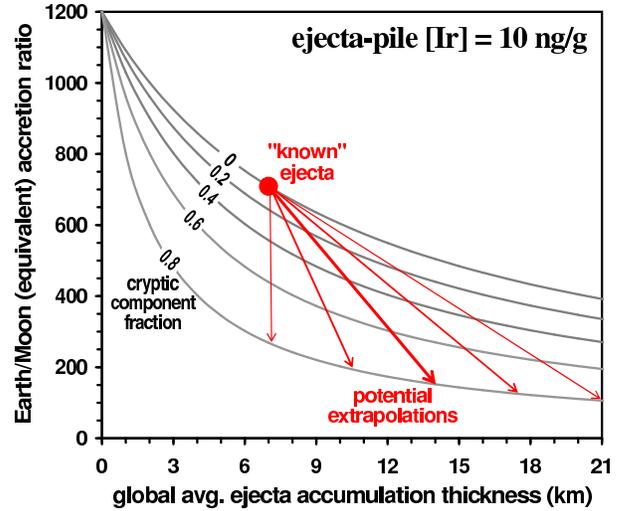, scale=0.45}}
\caption{Earth/Moon (equivalent) accretion ratio, shown as a function of
  assumed mean global thickness of the ejecta layer. The various curves are
  labeled to indicate different proportion of the cryptic HSE component as a
  fraction of the total lithospheric HSE component. The cryptic component
  represents the assumed sequestered HSE component and, in addition, accounts
  for never-accreted impactor matter that may have been delivered during
  the Late Heavy Bombardment. The ejecta layer is assumed to contain 10~ng/g
  Ir.}
\label{fig4}
\end{figure}

We therefore model the total impact mass that has collided with the lunar
 crust, what may be thought of as the equivalent veneer mass, as a combination
 of two main components: the ejecta layer itself, with $\sim 10~\rm{ng/g}$ Ir
 abundance, and an unobservable, cryptic component. This cryptic component
 represents the assumed sequestered component of metal that settled out as a
 result of larger impacts and, in addition, also accounts for never accreted
 impactor matter that may have been delivered during the Late Heavy
 Bombardment. The results in terms of the Earth to Moon (equivalent) accretion
 ratio, accounting for both the lunar mantle and crust components, are shown
 in Figure \ref{fig4}, where the various curves correspond to different
 assumptions for the proportion of the cryptic HSE component as a fraction of
 the total lithospheric HSE component. Even if we assume that the cryptic
 component is negligible and that no ejecta accumulation occurred beyond $\sim
 7~\rm{km}$ from `known' basins with diameters $\gtrsim 300~\rm{km}$, the
 Earth/Moon veneer mass ratio is $<700$. This result by itself implies that
 the total lithospheric HSE component is important, and probably larger than
 the mantle component, unless the sunken-sequestered component is very
 small. More realistically, the sequestered HSE fraction is probably $\sim
 0.5$, but anything between 1/5 and 2/3 seems almost equally plausible, and
 the total ejecta accumulation is likely at least 2 times the `known' ejecta
 volume. The extrapolated Earth/Moon veneer mass ratio is then roughly 300,
 assuming 10 ng/g Ir in the ejecta layer. This ratio would be even lower, if
 inefficient accretion during the Late Heavy Bombardment contributed
 significantly to the cryptic component. The huge uncertainties in the
 sequestered HSE fraction and the contribution of inefficient accretion to the
 cryptic component are permissive of the Earth/Moon veneer mass ratio being
 conceivably as low as $\sim 150$ or as high as $\sim 700$.

The implied total masses of chondritic-debris additions associated with our
various models are perhaps most easily comprehended when expressed in terms of
equivalent proportion of chondritic matter within the lunar crust (the
`equivalent' qualifier is needed because we assume that a major fraction of
the HSE actually became sequestered at the bottoms of the deepest impact melt
pools near the base of the crust). Assuming the mass of the crust is $5 \times
10^{24}~\rm{g}$, models that suggest an Earth/Moon accretion ratios of 600,
300 and 150, imply equivalent proportions of chondritic matter within the
crust of about 0.34\%, 1.0\% and 2.4\% by weight, respectively \footnote{For
this illustrative calculation, we assume that all the material delivered to the
Earth and Moon was accreted.}. These proportions represent additions to the
$1.7 \times 10^{22}~\rm{g}$ of chondritic matter inferred to be present in the
mantle source region of the lunar (mare) basalts.

In summary, we find that the Earth/Moon mass accretion ratio likely ranges
from $\sim 150$ to $\sim 700$. In addition, we note that comparison of the
mass accretion estimates for the lunar crust and mantle suggests that most of
the late veneer was deposited into the lunar crust rather than mantle, which
indicates that most of the late veneer was accreted by the Moon after the
formation of a permanent lunar crust \citep{WH04}, i.e. about 100~Myrs after
the Moon-forming impact.

\section{Planetesimal accretion with Gravitational Focusing}
\subsection{The Accretion of Leftover Planetesimals by the Earth and Mars}
\begin{figure}[htp]
\centerline{
\epsfig{file=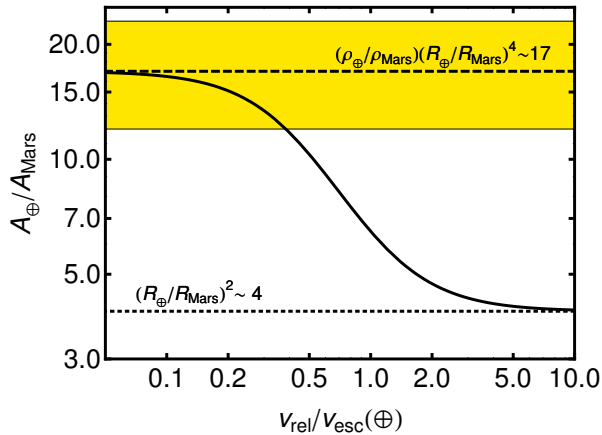, scale=0.7}}
\caption{Ratio of the mass accretion rates for the Earth and Mars,
$A_{\earth}/A_{Mars}$, as a function of $v_{rel}$ in units of Earth's
escape velocity, $v_{esc}(\Earth)$. The dotted line represents the limit
without any gravitational focusing ($v_{rel} \gg v_{esc}(\Earth)$), in which
case $A_{\Earth}/A_{Mars}=(R_{\Earth}/R_{Mars})^2 \sim 4$. The dashed line
corresponds to the limit with strong gravitational focusing ($v_{rel} \ll
v_{esc}(\Earth)$), in which case
$A_{\Earth}/A_{Mars}=(\rho_{\Earth}/\rho_{Mars})(R_{\Earth}/R_{Mars})^4 \sim
17$. The yellow shaded region shows the range of Earth/Mars mass accretion
ratios that are consistent with the inferred late veneer for the Earth and
Mars.}
\label{fig3}
\end{figure}

If the relative velocity between the planetesimal and the terrestrial planets, $v_{rel}$, is less than the escape velocity from the terrestrial planets
then the accretion cross sections of the terrestrial planets are enhanced above
their geometric values by gravitational focusing. The gravitationally enhanced
cross section is given by
\begin{equation}\label{e0}
A=\pi R^2 \left[1+\left(\frac{v_{esc}}{v_{rel}}\right)^2\right].
\end{equation}
Using Equation \ref{e0}, we can write the accretion cross section ratio of
Earth/Mars as
\begin{equation}\label{e1}
\frac{A_{\Earth}}{A_{Mars}}=\frac{(1+(v_{esc}(\Earth)/v_{rel})^2)R_{\Earth}^2}{(1+(v_{esc}(Mars)/v_{rel})^2)R_{Mars}^2},
\end{equation} 
where the subscripts $\Earth$ and `Mars' label the quantities corresponding to
Earth and Mars, respectively. If $v_{rel} \ll v_{esc}$, Equation \ref{e1} can
be simplified to
\begin{equation}
\frac{A_{\Earth}}{A_{Mars}}\sim
\frac{\rho_{\Earth}}{\rho_{Mars}}\left(\frac{R_{\Earth}}{R_{Mars}}\right)^4
\sim 17,
\end{equation} 
where $\rho_{\Earth}$ and $\rho_{Mars}$ correspond to the mean density of the
Earth and Mars, respectively. This implies that the accretion ratio between
the Earth and Mars that has been estimated from the HSE abundances
($A_{\Earth}/A_{Mars} \sim 12-23$, see section 3) agrees very well with the
expected mass accretion ratio between the Earth and Mars, if the accretion
cross sections of the Earth and Mars were significantly enhanced by
gravitational focusing. For comparison, if $v_{rel}$ had been larger than the
escape velocities of Earth and Mars such that gravitational focusing becomes
irrelevant, then the ratio of the accretion cross section Earth/Mars is simply
given by $(R_{\Earth}/R_{Mars})^2 \sim 4$. This value is significantly lower
than the mass ratio of the late veneer that has been estimated from the
terrestrial and Martian abundances of HSEs (see Figure \ref{fig3}). We
therefore conclude that we can account for the relative amounts of late veneer
accreted by the Earth and Mars, if it was delivered concurrently by a
population of planetesimals with a velocity dispersion small enough such that
$v_{rel} \ll v_{esc}$.

\subsection{The Accretion of Leftover Planetesimals by the Earth and the
  Moon} 
\begin{figure}[htp]
\centerline{
\epsfig{file=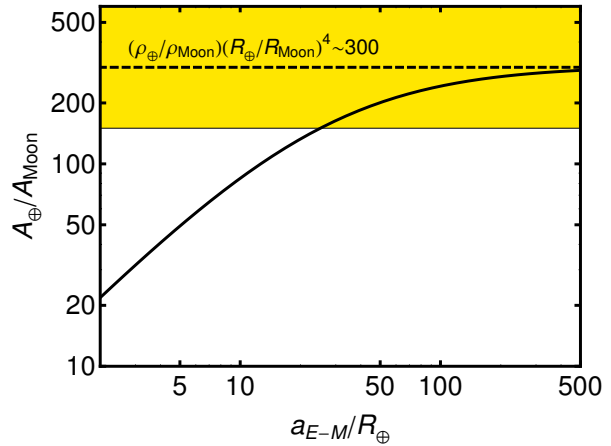, scale=0.7}}
\caption{Ratio of the mass accretion rates for the Earth and Moon,
$A_{\earth}/A_{Moon}$, as a function of the Earth-Moon distance in Earth radii,
$a_{E-M}/R_{\earth}$ with strong gravitational focusing (i.e., $v_{rel} \ll
v_{esc}$). The solid line is the analytic result from \citet{BS73} given in
Equation (\ref{e7}). The dashed line represents our estimate for the ratio of
the Earth/Moon accretion cross section from Equation \ref{e2}, which is valid
for large Earth-Moon separations when the Earth and the Moon can be well
approximated as isolated bodies. The yellow shaded region shows the range of
Earth/Moon mass accretion ratios that are consistent with the inferred late
veneer added to the Earth and Moon based on HSE observations.}
\label{fig1}
\end{figure}

We can extend the above argument to the Earth and the Moon to get a
rough estimate for the Earth/Moon accretion ratio. However, using the
expression for gravitational focusing from Equation \ref{e1} is strictly
speaking only valid for isolated bodies and is therefore only a rough
approximation for the Earth/Moon accretion ratio. Estimating the
Earth/Moon accretion ratio from Equation \ref{e1} we have
\begin{equation}\label{e2}
\frac{A_{\Earth}}{A_{Moon}}\sim
\frac{\rho_{\Earth}}{\rho_{Moon}}\left(\frac{R_{\Earth}}{R_{Moon}}\right)^4
\sim 300.
\end{equation}
This implies that in the limit in which we can treat the Earth and Moon as
isolated bodies, i.e., for very large Earth-Moon separations, the Earth should
have accreted about 300 times more mass as late veneer compared to the Moon
(see Figure \ref{fig1}).

\begin{figure}[htp]
\centerline{
\epsfig{file=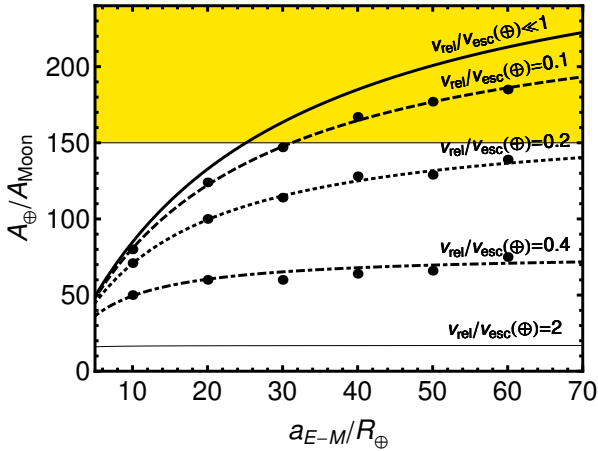, scale=0.7}}
\caption{Ratio of the mass accretion rates for the Earth and Moon,
$A_{\earth}/A_{Moon}$, as a function of the Earth-Moon distance in Earth radii,
$a_{E-M}/R_{\Earth}$ for various relative velocities, $v_{rel}$, from Equation
(\ref{e7}) \citep{BS73}. For comparison, we also show the results from our
numerical simulations (shown as points) in which we integrate the trajectories
of the planetesimals in the Earth-Moon system directly. The yellow shaded
region shows the range of Earth/Moon mass accretion ratios that are consistent
with the inferred terrestrial and lunar late veneer based on HSE
observations.}
\label{fig2}
\end{figure}

\citet{BS73} derived analytically the ratio of the Earth/Moon accretion cross
section. Assuming an isotropic planetesimal velocity distribution far from the
Earth and neglecting Earth's shadow they find
\begin{equation}\label{e7}
\frac{A_{\Earth}}{A_{Moon}}\geq
\frac{1+(v_{rel}/v_{esc}(\Earth))^2}{\frac{7}{6}(R_{\Earth}/a_{E-M})+0.045+(v_{rel}/v_{esc}(\Earth))^2}\left(\frac{R_{\Earth}}{R_{Moon}}\right)^2
\end{equation} 
where $a_{E-M}$ is the Earth-Moon separation. Equation (\ref{e7}) is a lower
limit to $A_{\Earth}/A_{Moon}$ because the effect of Earth's shadow was
neglected in deriving the accretion cross section for the Moon, i.e., it
neglects lunar impactors that would have collided with Earth first. We
confirmed this analytic result for the Earth/Moon accretion cross section by
direct numerical integrations of planetesimal trajectories in the Earth-Moon
system (see Figure \ref{fig2}).

Figure \ref{fig1} shows the Earth/Moon accretion ratio as a function of the
Earth-Moon separation for $v_{rel} \ll v_{esc}$. As expected the Earth-Moon
accretion cross section approaches the ratio for isolated bodies at large
Earth-Moon separations. At the current Earth-Moon separation, which
corresponds to about $60~\rm{R_{\Earth}}$, the accretion ratio is $\sim
200$. The Earth/Moon accretion ratio increases with increasing Earth-Moon
separation approaching the limit derived in Equation \ref{e2} for isolated
bodies, this behavior remains unchanged as long as gravitational focusing
plays a significant role in enhancing the accretion cross section above the
geometric value (see Figure \ref{fig2}). This result may seem surprising at
first, because Earth's gravitational field accelerates the incoming
planetesimals such that the Moon's gravitational focusing is reduced. However,
Earth's gravitational field also focuses the incoming planetesimals such that
the Moon intercepts a larger number of planetesimals than it would have
otherwise.  As a result, the Earth/Moon accretion cross section decreases with
decreasing Earth-Moon separation, which implies that it was lower in the past
before the Moon evolved tidally outward to its current location. The tidal
evolution timescale for the Moon to evolve from an initial separation of a few
times Earth's radius to a separation $a_{E-M} \gg R_{\earth}$ is
\begin{equation}\label{e001}
t_{tidal}= \frac{2}{39}\frac{Q}{k}
\frac{M_{\earth}}{M_{Moon}}\left(\frac{a_{E-M}}{R_{\earth}}\right)^5
\left(\frac{GM_{\earth}}{a_{E-M}^3}\right)^{-1/2}\\
 \sim 1.3 \times 10^4 
\frac{Q/12}{k/0.299}\left(\frac{a_{E-M}}{10R_{\Earth}}\right)^{6.5}~\rm{yr}
\end{equation}
where $k$ and $Q$ are the tidal dissipation function and the tidal Love number
of the Earth, respectively. From Equation \ref{e001} we see that the initial
tidal evolution was very fast such that the Earth-Moon system did not spend a
significant amount of time at small Earth-Moon separations and evolved to
$a_{E-M} \gtrsim 40 R_{\Earth}$ within about 110~Myrs of the Moon forming
impact. This estimates assumes that $Q\sim 12$ and that its value did not
significantly change throughout the tidal evolution. In section 6 we calculate
a planetesimal accretion timescale of $\sim 170~\rm{Myr}$, which suggests that
most of the planetesimals were accreted after the Earth-Moon system evolved to
separations $\gtrsim 40 R_{\earth}$.  We therefore conclude that the relative
amounts of late veneer added to the Moon and Earth, as inferred from their HSE
abundances, are consistent with the accretion of small planetesimals with a
velocity dispersion $u \lesssim v_{rel} \sim 0.1 v_{esc}({\Earth})$. In
contrast, if $v_{rel} \gg v_{esc}$, i.e. in the limit without any
gravitational focusing, the Earth/Moon accretion ratio is independent of the
Earth-Moon separation and is simply given by the ratio of the geometric cross
sections $(R_{\Earth}/R_{Moon})^2 \sim 14$ (see Figure \ref{fig2}), which is
inconsistent with the Earth/Moon mass accretion ratio inferred from HSEs of
150-700.

\section{Planetesimal Sizes} 
We have shown in section 4 that the relative amounts of late veneer delivered
to Earth, Moon and Mars can be explained by the accretion of small
planetesimals after giant impacts provided that gravitational focusing played
a significant role in increasing the accretion cross section above the
geometric value. Since gravitational focusing only acts when $v_{rel} \ll
v_{esc}$, we conclude that the velocity dispersion of the planetesimals was
comparable to or less than the velocity dispersion of the terrestrial planets,
i.e. $u \lesssim v$, such that $v_{rel} \sim v$. Further, we can use our
estimate from section 2 that $v \sim 0.1 v_{esc}$ to place an upper limit on
the typical planetesimal size as follows.

The velocity dispersion of the planetesimals is stirred gravitationally by the
terrestrial planets and damped by mutual planetesimal collisions such that
\begin{equation}
\frac{1}{u}\frac{du}{dt} \sim \frac{\Sigma \Omega}{\rho R}
\left(\frac{v_{esc}}{v_{rel}} \right)^2 \left(\frac{v_{esc}}{u} \right)^2 -
\frac{\sigma \Omega}{\rho_s s}
\end{equation}
where $s$ and $\rho_s$ are the radius and density of the planetesimals and
$u \leq v_{rel}$. Balancing the stirring and the damping rates yields
\begin{equation}\label{e5}
s \sim \frac{\sigma \rho}{\Sigma \rho_s}
\left(\frac{v_{rel}}{v_{esc}}\right)^2 \left(\frac{u}{v_{esc}}\right)^2 R.
\end{equation} 
Evaluating Equation (\ref{e5}) for $\sigma/\Sigma \sim 1\%$ (see
discussion in section 2), substituting for $v_{rel}\sim v \sim 0.1 v_{esc}$
and using $u \sim v$, we have
\begin{equation}\label{e6}
s \sim 10~\rm{m}
\end{equation}
where we assumed $\rho_s \sim 3~\rm{g/cm^3}$ and $R \sim R_{\Earth}$ and $\rho
\sim \rho_{\Earth}$. This implies that $u \lesssim v \sim 0.1 v_{esc}(\earth)$
as long as the typical planetesimals, that damped the eccentricities and
inclinations of the terrestrial planets and that delivered the late veneer,
were smaller than about 10 meters in size. We assumed when evaluating Equation
(\ref{e6}) that $u \sim v$. The actual planetesimal sizes could therefore be
smaller than estimated in Equation (\ref{e6}), which would imply that $u <
v$. The corresponding optical depth of such a planetesimal population is
$\tau \sim \sigma/\rho_s s$, which implies $\tau \gtrsim 2.5 \times
10^{-5}$. We also note here that recent work by \citet{W11} showed that the
size distribution of the asteroid belt can be reproduced by coagulation from
an initial population of planetesimals as long as they have sizes $\lesssim
100~\rm{m}$, providing independent support for a planetesimal population in
the inner solar system that was smaller than about 100 meters in size.

If typical planetesimal sizes would have exceeded about 10 meters, then $u>v$
such that $v_{rel} \sim u >0.1 v_{esc}$, which would imply weaker or no
gravitational focusing, making it hard to reconcile the resulting Earth/Moon
and Earth/Mars accretion ratios with the relative quantities of late veneer
delivered to these bodies. Furthermore, if $u>v \sim 0.1 v_{esc}$ than this
would imply that more mass, than we estimated in section 2, must have been
residing in small planetesimals at the end of giant impacts in order to damp
the eccentricities and inclinations of the terrestrial planets.

\section{Gap Formation and Accretion Timescales}
Gaps were not important before the end of the giant impact phase, because the
radial separation of protoplanets was only a few times larger than the widths
of their Hill radii. But at the end of giant impacts, when the terrestrial
planets achieved large-scale orbital stability, their orbital separation was
much larger than their Hill radii and gaps likely formed around their orbits
\citep{GLS042}. Gaps increase the clean up timescale of the leftover
planetesimals, because accretion onto the protoplanets can now only proceed
from the gap edges. 
Following \citet{GLS042}, the rate at which
a terrestrial planet accretes small planetesimals from gap edges located a
distance x from the planets' semi-major axis is
\begin{equation}
\frac{1}{M}\frac{dM}{dt} \sim \frac{\sigma_0 \Omega}{\rho R}
\left(\frac{2x}{\Delta a}\right)^4 \left(\frac{v_{esc}}{v_{rel}}\right)^2
\end{equation} 
where $\Delta a$ is the distance between neighboring planets. The gap
surface density of the small bodies follows a power law such that the mass
surface density at the gap edges is given by $\sigma =\sigma_0
(2x/\Delta a)^4$. Writing $t_{acc}=-\sigma_0 \frac{dt}{d\sigma_0}
\sim 2 \pi \sigma_0 a \Delta a (dM/dt)^{-1}$ yields:
\begin{equation}\label{ex1}
t_{acc}=-\sigma_0 \frac{dt}{d\sigma_0} \sim \frac{\rho R}{\Sigma
  \Omega}\left(\frac{v_{rel}}{v_{esc}}\right)^2 \left(\frac{\Delta
  a}{2x}\right)^4
\end{equation}
where $\Sigma = M/(2\pi a \Delta a)$. The spacing between the terrestrial
planets is $\Delta a \sim a/3$ and, since $v>u$, $x$ will be roughly
given by the radial excursions of the planet which is $\sim ae$. This yields a
clean up timescale of
\begin{equation}
t_{acc}=-\sigma_0 \frac{dt}{d\sigma_0} \sim \frac{\rho R}{\Sigma
  \Omega}\left(\frac{v_{rel}}{v_{esc}}\right)^2 \left(\frac{1}{6e}\right)^4
\sim 170~\rm{Myrs}
\end{equation}
when evaluated at 1~AU for $R=R_{\earth}$, $M=M_{\earth}$ and $v_{rel} \sim
ea\Omega \sim 0.1 v_{esc}$. This implies that most of the planetesimals were
accreted after the formation of a permanent lunar crust. This result is
consistent with the fact that the majority of the late veneer seems to reside
in the lunar crust rather than mantle (see section 3 for details). Because the
planetesimal accretion timescale is long compared to the tidal evolution
timescale of the Earth-Moon system (see Equation (10) in section 4), this
implies that the Earth-Moon separation was already $\gtrsim 40~R_{\earth}$
when most of the late veneer was delivered to the Earth-Moon system.

\section{Discussion and Conclusions}
The abundances of HSEs suggest that a total of about $0.01~M_{\earth}$ was
delivered as `late veneer' by planetesimals to the terrestrial planets after
the giant impact phase. We showed here that small residual planetesimals, with
radii $\lesssim 10~\rm{m}$, containing about $\sim 1\%$ of the mass of the
Earth can provide the dynamical friction needed to relax the terrestrial
planets' eccentricities and inclinations after giant impacts and
simultaneously account for the relative and absolute amounts of late veneer
added to Earth, Moon and Mars. Small planetesimal sizes are required to ensure
efficient damping of the planetesimal's velocity dispersion by mutual
collisions, which in turn ensures sufficiently low relative velocities between
the terrestrial planets and the planetesimals such that the planets' accretion
cross sections are significantly enhanced by gravitational focusing above
their geometric values. Specifically we find that, if $v_{rel} \ll v_{esc}$,
gravitational focusing yields a mass accretion ratio Earth/Mars $\sim
(\rho_{\earth}/\rho_{mars}) (R_{\earth}/R_{mars})^4 \sim 17$, which agrees
well with the mass accretion ratio inferred from HSEs of 12-23.  For the
Earth-Moon system, we find a mass accretion ratio of $\sim 200$, which, as we
show in section 3, is consistent with estimates of 150-700 derived from HSE
abundances that include the lunar crust as well as mantle component. Further,
we find that the higher abundance of siderophilic elements in the lunar crust
compared to the lunar mantle is consistent with the idea that most of the late
veneer was delivered by small planetesimals. This is because, if the residual
planetesimals were indeed small, gaps will likely form around the terrestrial
planets, which will prolong the planetesimal accretion timescale such that
most of the late veneer is added to the lunar crust after the lunar mantle was
isolated by the formation of a permanent crust. We note here that, although we
suggest that the majority of the accreted late veneer was delivered by small
planetesimals, larger planetesimals were certainly residing among the small
planetesimal population and must have played a crucial role the mixing and
settling of HSEs in the lunar crust.

\citet{BW11} suggested recently that most of the late veneer may have been
delivered by a few very large planetesimals with the largest terrestrial
impactor exceeding more than 1000~km in radius. Delivering the majority of the
late veneer by one or two very large bodies may explain stochastically a large
mass accretion ratio between the Earth and Moon. However, whereas small body
accretion can account for the relative quantities of late veneer inferred from
HSE abundances for the Earth/Moon and Earth/Mars simultaneously, it would
remain a coincidence in a stochastic accretion scenario. Further, if a small
number of 1000~km sized planetesimals were indeed responsible for the late
veneer, then these planetesimals cannot have damped the eccentricities and
inclinations of the terrestrial planets after giant impacts. This is because,
if most of the planetesimal mass resided in such large bodies, they would have
to have a significantly higher velocity dispersion, because mutual
planetesimal collisions, that damp their velocities are significantly less
frequent for larger planetesimals compared to small ones (see section 5). In
this case $v_{rel}$ would be determined by the velocity dispersion of these
large planetesimals, which in turn implies that significantly more than 1\% of
the total mass would be required in large planetesimals to damp the
eccentricities and inclinations of the terrestrial planets (see section
2). This, however, would be inconsistent with the $\lesssim 0.01~M_{\earth}$
of chondritic material delivered as late veneer to the Earth, Moon and
Mars. In principle a population of planetesimals made of primarily silicates
with extremely low HSE abundances could have provided the required dynamical
friction. However, the relative abundances of the different HSEs in the
terrestrial planets and the Moon are consistent with chondritic material and
hence favor the idea that they were delivered by small, undifferentiated
planetesimals with chondritic composition.

Finally, as we have shown in section 3, the Earth/Moon impact ratio
likely falls in the range 150-700, once the HSE deposited into the lunar crust
and the upper part of the lunar lithosphere are accounted for (see also
\citet{WH04}). If the ratio of the late veneer accreted by the Earth and Moon
falls at the lower end of this range then it is consistent with small body
accretion. However, if it can be conclusively shown that the Earth/Moon mass
accretion ratio lies at the upper end of this range then it cannot be
explained by the small body accretion discussed here. In this case it could
instead be either due to a small number of stochastic events that delivered
most of the late veneer \citep{BW11} or due to a smaller retention fraction of
the material delivered to the Moon compared to Earth as might be expected for
impact velocities significantly exceeding the lunar escape velocity.

\acknowledgements{We thank David Jewitt and our two referees, John Chambers and
  Richard Walker, for valuable comments that helped to improve this
  manuscript. For HS support for this work was provided by NASA through Hubble
  Fellowship Grant \# HST-HF-51281.01-A awarded by the Space Telescope Science
  Institute, which is operated by the Association of Universities for Research
  in Astronomy, Inc., for NASA, under contact NAS 5-26555. PW acknowledges
  support from NASA grant NNX09 AE31 G. QZY acknowledges NASA Cosmochemistry
  (NNX11AJ51G) and Origins of Solar Systems (NNX09AC93G) grants for support.}

\end{document}